\documentstyle[preprint,eqsecnum,aps]{revtex}

\begin{document} 
\draft 
\title{Apparent and Actual Shifts in Mass and Width of\\
       $\phi$ Mesons Produced in Heavy-Ion Collisions}
\author{R. S. Bhalerao}
\address{Nuclear Theory Center, Indiana University\\
2401 Milo B. Sampson Lane, Bloomington, IN 47405, USA\\
and\\
Theoretical Physics Group,
Tata Institute of Fundamental Research\\
Homi Bhabha Road, Colaba, Mumbai 400 005, India$^1$}
\author{S. K. Gupta}
\address{Nuclear Physics Division,
Bhabha Atomic Research Center\\
Trombay, Mumbai 400 085, India}
\maketitle
\begin{abstract}
We present a method of analyzing invariant-mass spectra of kaon pairs
resulting from decay of $\phi$ mesons produced in high-energy
heavy-ion collisions. It can be used to extract the shifts in the mass
and the width ($\Delta M$ and $\Delta \Gamma$) of the $\phi$ mesons
when they are inside the dense matter formed in these collisions. We
illustrate our method with the help of available preliminary data.
Extracted values of $\Delta M$ and $\Delta \Gamma$ are significantly
larger than those obtained with an earlier method. Our results are
consistent with the experimentally observed $p_T$ dependence of the
mass shift. Finally, we present a phenomenological relation between
$\Delta M$ and $\Delta \Gamma$. It provides a useful constraint on
theories which predict the values of these two quantities.
\end{abstract}

\newpage


\vfill

\noindent{$^1$ permanent address}

\noindent{E-mail: bhalerao@theory.tifr.res.in, ~~~~~~~~~~ 
Fax: 091 22 215 2110}

\noindent{E-mail: skgupta@magnum.barct1.ernet.in, ~~Fax: 091 22 556 0750}

\newpage
\noindent{\bf {1~ Introduction}}
\medskip

Properties such as mass, decay width, size, etc. of hadrons in general
and light vector mesons ($\rho, \omega, \phi$) in particular, are
thought to undergo a change when the surrounding medium is raised to a
higher temperature and/or density; for recent reviews of
finite-temperature and finite-density effects, see Refs. \cite{hat1} and
\cite{hat2}, respectively. Interest in this scenario arises largely
due to its close connection with phase transitions in quantum
chromodynamics (QCD). Recent years have witnessed intense experimental
and theoretical activity in this area \cite{qm95}. In this paper, we
focus on the mass as well as the width of the $\phi$ meson.

Recently E-802 experimental collaboration at AGS (BNL) has reported
preliminary results on the shift in the mass of the $\phi$ mesons
produced in central $Si + Au$ collisions at 14.6 A GeV/c
\cite{akiba95,wang95}. The mass was determined from the 
invariant-mass distribution of kaon pairs arising from the dominant
decay mode $\phi \rightarrow K^+ K^-$. It was studied as a function of
the multiplicity of the collision events and the transverse momentum
$p_T$ of the accepted $\phi$'s. It was found that in the events with
the highest multiplicity (top 2$\%$ target-multiplicity-array TMA cut)
the mass drops by $2.3 \pm 0.9 \pm 0.1$ MeV compared to the free-space
value $1019.413 \pm 0.008$ MeV \cite{pdg}. For the $\phi$ mesons with
$p_T < 1.25$ GeV/c, the shift was even more ($3.3 \pm 1.0 \pm 0.1$
MeV), whereas for $p_T > 1.25$ GeV/c, there was no apparent shift. No
numbers were reported for the shift in the {\it decay width} of the
$\phi$ mesons. However, from the confidence contours for 1, 2 and 3
standard deviations in the observed mass versus width plot given in
\cite{wang95}, it appears that the central value of the width is higher
than the free-space value $4.43 \pm 0.06$ MeV \cite{pdg} by about 0.78
MeV. {\it If confirmed, this will be the first evidence of the
modification of hadronic properties inside dense matter formed in
heavy-ion collisions.}

In a more recent publication \cite{akiba96}, the E-802 collaboration has
reported no shift in the mass and the width of the $\phi$ meson. This,
however, does not contradict the earlier observations \cite{wang95}
because in Ref. \cite{akiba96} the top 7$\%$ of the highest-multiplicity
events were clubbed together and no multiplicity dependence of the
mass was reported. In the earlier work \cite{wang95}, the shift in the
mass was seen when the TMA cut was applied at 2$\%$.

\newpage
\noindent{\bf {2~ Method}}
\medskip

In this paper, we present a different method of analyzing the
invariant-mass spectra of kaon pairs. We then use the data in
Refs. [4-5] only to illustrate our method. We find that our method
yields shifts in the $\phi$-meson mass and width which are several
times larger than the values quoted above. We caution the reader that
since these data are preliminary, our numerical results are obviously
subject to change. {\it However, our main point is the method
presented here which is independent of whether the published data
[4-5] are eventually confirmed or not.} We think the present method
has relevance to the analysis of {\it future data} on $\phi$ production.
This acquires added importance in view of the possibility of a similar
experiment being performed at CERN APS \cite{miake}.

Let $M_0$ and $M_1$ be the rest masses and let $\Gamma_0$ and $\Gamma_1$
be the widths of the $\phi$ meson in the free space and in the dense
medium, respectively. We define the shifts as
\begin{equation}
\Delta M = M_0 - M_1, ~~~\Delta \Gamma = \Gamma_1 -\Gamma_0,
\eqnum{1}
\end{equation}
so that both are positive when the mass drops and the width increases
with respect to their free-space values.

In Ref. \cite{wang95}, the invariant-mass spectrum of the $K^+K^-$
pairs was fitted by a function consisting of a background term and a
relativistic Breit-Wigner (BW) resonant term convoluted with a
Gaussian experimental mass resolution function. This procedure yielded
the values of the shifts $\Delta M$ and $\Delta \Gamma$ given
above. A background-subtracted mass spectrum is displayed in Fig. 1
where the histogram corresponds to 3$\%$ TMA and low-$p_T$ cuts and is
taken from Fig. 2(b) of Ref. \cite{wang95}. The dashed and solid
curves represent single BW resonance terms convoluted with a Gaussian
as stated above. Areas under the two curves are normalized to that
under the histogram. The dashed curve corresponds to the scenario
where there is no shift in the mass and the width. The solid curve
corresponds to $\Delta M = 2.3$ MeV and $\Delta \Gamma = 0.78$
MeV. Clearly the data indicate a shift in the mass and the width of
the $\phi$ meson. However, are the values of $\Delta M$ and $\Delta
\Gamma$ obtained by fitting a {\it single} BW resonance term to the
(background-subtracted) data correct? We think they are not.

Since the mean lifetime of $\phi$ in its rest frame is about 45 fm/c,
a majority of $\phi$'s are expected to decay long after the dense
medium in which they were produced has ceased to exist. That is, they
will decay essentially in free space ($\Delta M = 0 = \Delta
\Gamma$). The rest of the $\phi$'s, however, decay while still inside
the dense medium. Hence a better procedure would be to fit the
background-subtracted data with two instead of one BW terms, one
unshifted and another shifted, added with appropriate weights. This we
now proceed to do.

We work in the center-of-mass frame of the dense medium formed in the
nucleus-nucleus collision. We employ the natural units $c=\hbar=1$.
Let $f$ be the fraction of $\phi$'s decaying {\it inside} the medium;
then $(1-f)$ is the fraction decaying in free space. We reanalyze the
background-subtracted data of Ref. \cite{wang95} by fitting them with
two BW terms --- an unshifted BW with mass $M_0$, width $\Gamma_0$ and
weight $(1-f)$ and a shifted BW with 
mass $M_1$, width $\Gamma_1$ and weight $f$:
\begin{equation}
{dN_{K^+K^-} \over dM} = (1-f)~ BW_c(M, M_0, \Gamma_0) 
+ f~ BW_c(M, M_1, \Gamma_1).
\eqnum{2}
\end{equation}
Here $BW_c$ denotes the relativistic Breit-Wigner resonance term
convoluted with a Gaussian experimental mass resolution function:
\begin{eqnarray}
BW_c(M, M_0, \Gamma_0) &=& \int BW(M', M_0, \Gamma_0) \exp\left[-{1 \over 2}
\left({M-M' \over \sigma}\right)^2\right] {dM' \over \sigma \sqrt{2 \pi}},
\eqnum{3} \\
BW(M, M_0, \Gamma_0) &=& {M_0~ \Gamma_0(M) \over \pi} {2M \over (M^2
 - M_0^2)^2 + M_0^2 ~\Gamma_0^2(M)}. \eqnum{4}
\end{eqnarray}
The experimental mass resolution $\sigma$ is taken to be 2.2 MeV 
\cite{wang93}. The energy-dependent width $\Gamma_0(M)$ is taken to be
\cite{jackson}
\begin{equation}
\Gamma_0(M) = \Gamma_0 \left({q \over q_0}\right)^3 {2 q_0^2 \over q^2 
+ q_0^2}, \eqnum{5}
\end{equation}
where the kaon momenta $q$ and $q_0$ are given by
\begin{equation}
q = (M^2/4 - M_K^2)^{1/2} ~~{\rm and}~~ q_0 = (M_0^2/4 - M_K^2)^{1/2}.
\eqnum{6}
\end{equation}
Expressions for $BW_c(M, M_1, \Gamma_1)$, $BW(M, M_1, \Gamma_1)$
and $\Gamma_1(M)$ are obtained by replacing $M_0$ by $M_1$ and
$\Gamma_0$ by $\Gamma_1$ in Eqs. (3-6).
We find that the numerical results with energy-dependent and
energy-independent widths to be practically the same.

We now derive an expression for $f$. Let $\tau_0 = \Gamma_0^{-1}$ and
$\tau_1 = \Gamma_1^{-1}$ be the mean lifetimes of $\phi$ {\it at rest} in
the free space and in the dense medium, respectively. Then the
corresponding lifetimes when the $\phi$ is {\it in motion} would be
$\tau_0 \gamma$ and $\tau_1 \gamma$, due to the time dilation. Here
$\gamma \equiv (1-v^2)^{-1/2}$ is the usual Lorentz factor and $v$ is
the average velocity of the $\phi$'s in the medium. Let $d$ be the
``size'' or ``radius'' or a typical linear dimension associated with
the extent of the medium. The average time the $\phi$'s produced in
the medium would take to traverse this distance is $d/v$. If $N_0$ is
the number of $\phi$'s at time $t=0$, then at time $t=d/v$, only $N_0
\exp(-\Gamma_1 d/v\gamma)$ of them would be left. Hence
$f=1-\exp(-\Gamma_1 d/v\gamma)$. Now
\[
v\gamma=p/M_1=\sqrt{E^2-M_1^2}/M_1=\sqrt{M_T^2~ {\rm cosh}^2y_{cm} 
- M_1^2}/M_1,
\]
where $p, E, M_T$ and $y_{cm}$ denote, respectively, the momentum, energy,
transverse mass and center-of-mass rapidity of the $\phi$ traversing the
medium. Hence 
\begin{equation}
f(\Delta M, \Delta \Gamma) = 1 -\exp(-M_1 \Gamma_1 d/\sqrt{M_T^2~ {\rm
cosh}^2y_{cm} - M_1^2}). \eqnum{7}
\end{equation}

Interactions of the outgoing $K^\pm$ with the medium were ignored in
Refs. [4-5] and we too shall ignore them. This is a good first
approximation because if these interactions were important the $\phi$
peak would have been washed out, whereas experimentally, a distinct,
narrow peak is seen. Given the quality of the available data, we think,
the present approach is adequate. In view of the possibility of
better-quality data becoming available in the future, it is desirable
to include these interactions by performing a detailed Monte-Carlo
calculation. This would entail a considerable amount of extra work,
and we leave it to the future. Such an elaborate calculation may
change the values of $\Delta M$ and $\Delta \Gamma$ to some extent,
but we do not expect it to change our conclusion that these will be
more realistic than those resulting from the one-BW fit to the
data. If the kaon-medium interaction is ignored, then $M_T$ and
$y_{cm}$ of the $\phi$ are the same as those of the $K^\pm$ pair. In a
Monte-Carlo simulation one can extract them from experimental data by
modeling the interaction.

\newpage
\noindent{\bf {3~ Results and discussion}}
\medskip

Results of a least-squares fit to the same experimental data as in
Fig. 1, with two instead of one BW terms, are shown in Fig. 2. The two
dashed curves in Fig. 2 correspond to the two convoluted BW terms, one
shifted and the other unshifted. The solid curve corresponds to their
weighted sum as in Eqs. (2-6). The input parameters are $d = 5$ fm,
$M_T = 1100$ MeV and $y_{cm} = 0.3$ \cite{akiba95}. The two fitted
parameters are $\Delta M = 6.0 \pm 1.9 $ MeV and $\Delta \Gamma = 5.4
\pm 3.9 $ MeV. The resultant value of $f$ is 0.37. If $M_T = 1200$
MeV, the two fitted parameters are $\Delta M = 6.2 \pm 2.4 $ MeV and
$\Delta \Gamma = 7.7 \pm 5.0 $ MeV. The resultant value of $f$ is
0.34. Note that the above values of the shifts are several times
larger than those obtained with a one-BW fit. Secondly, even for an
unrealistically large value of $d$, say 10 fm, the fraction $f$ of
$\phi$'s decaying inside the dense matter is $\sim 0.4$ to 0.5 showing
that the assumption of a single BW resonance is questionable.

It is evident from Eq. (7) that as $p_T$ increases the fraction $f$
decreases. This means a larger fraction of $\phi$'s decay in free
space and hence the mass shift decreases. This is exactly what has
been observed experimentally \cite{wang95}.

Finally, we present an empirical relation between $\Delta M$ and 
$\Delta \Gamma$. We define
\begin{eqnarray}
\overline M &=& (1-f) M_0  + f M_1, \eqnum{8} \\
\Delta \overline M &=& M_0 - \overline M. \eqnum{9}
\end{eqnarray}
On substituting Eqs. (1) and (9) in Eq. (8) and simplifying we get
\begin{equation}
\Delta \overline M = f(\Delta M, \Delta \Gamma) \Delta M.
\eqnum{10}
\end{equation}
Given the experimental histogram as in Fig. 1, it is straightforward
to determine its centroid and it is reasonable to approximate
$\overline M$ by the centroid. Equation (9) can then be used to
determine the ``apparent'' shift 
$\Delta \overline M$. For a fixed $\Delta \overline M$,
Eq. (10) yields a curve $\Delta \Gamma$ versus $\Delta M$. We
illustrate this in Fig. 3 for $\Delta \overline M =$ 1, 2, 3 and 4
MeV. The input parameters are as above, namely $d = 5$ fm,
$M_T = 1200$ MeV and $y_{cm} = 0.3$.
Interestingly, as $\Delta M$ increases, $\Delta \Gamma$ decreases.
This is easy to understand from Eq. (10) because for a fixed $\Delta
\overline M$, as $\Delta M$ increases, $f$ has to decrease, which
from Eq. (7) 
requires $\Gamma_1$ and hence $\Delta \Gamma$ to decrease. The physics
of this is also clear if one considers the three curves in Fig. 2.
Equation (10) together with the experimentally determined $\Delta
\overline M$ provides a useful constraint on theories which predict
the values of $\Delta M$ and $\Delta \Gamma$. For example, if the
apparent shift $\Delta \overline M$ extracted from experimental data
is 2 MeV, then any model which claims to explain the data should have
$\Delta M$ and $\Delta \Gamma$ not inconsistent with the curve
labeled ``2'' in Fig. 3.

A variety of approaches, namely lattice QCD, QCD sum rules,
Nambu--Jona-Lasinio model, quantum hadrodynamics, bag models,
instanton-liquid model, etc. have been used in the literature to
predict the masses and widths of the light vector mesons at finite
temperatures ($T$) and/or densities ($\rho$). Here we focus on some of
the recent calculations for the $\phi$ meson.
Asakawa and Ko \cite{asakawa} have used the QCD sum rules to calculate
the $\phi$-meson mass when both $T$ and $\rho$ are finite. For the
values of $T$ and $\rho$ considered by them, the predicted drop in the
mass is as high as a few hundred MeV. They have proposed a double-peak
structure in the invariant-mass spectra of dilepton pairs arising from
$\phi$ decay as a signal of the transition from quark-gluon plasma to
hadronic matter.
Finite-temperature effects on the $\phi$-meson mass and width have
been studied recently by Shuryak and Thorsson, Haglin and Gale,
Bhattacharyya et al. and Song \cite{shuryak}.

\bigskip
\noindent{\bf {4~ Conclusions}}
\medskip

In conclusion, ({\it i}) we have presented a method of analyzing
invariant-mass spectrum of the kaon pairs; the resultant values of the
shifts in the $\phi$-meson mass and width are significantly larger
than those obtained with an earlier method. This difference arises due
to the long lifetime of the $\phi$ which ensures that only a small
fraction of them decay inside the dense matter. ({\it ii}) The model
presented here is consistent with the experimentally observed $p_T$
dependence of the mass shift. ({\it iii}) We have presented a
phenomenological relation between the shifts in the mass and the
width.

It is worth reiterating that the data in \cite{wang95} are
preliminary. In view of the interesting conclusions that could be
drawn from the data, it would be useful to have a more thorough
experimental investigation of the shift in the mass as well as the
width of the $\phi$ meson. One would like to have data with better
statistics, on a variety of targets and beams, and at various
energies. One would like to see invariant-mass spectra of not only the
kaon pairs but also the lepton pairs resulting from $\phi$ decay.

\acknowledgments

We thank Prof. K. V. L. Sarma for a critical reading of
the manuscript. We thank the organizers of the workshop on {\it
Quark-Gluon Plasma and Phase Transitions in the Early Universe}, Puri,
India, December 1995, during which this work was conceived and
started. 
One of us (RSB) would like to thank the Nuclear Theory Center,
Indiana University (USA) for the hospitality.

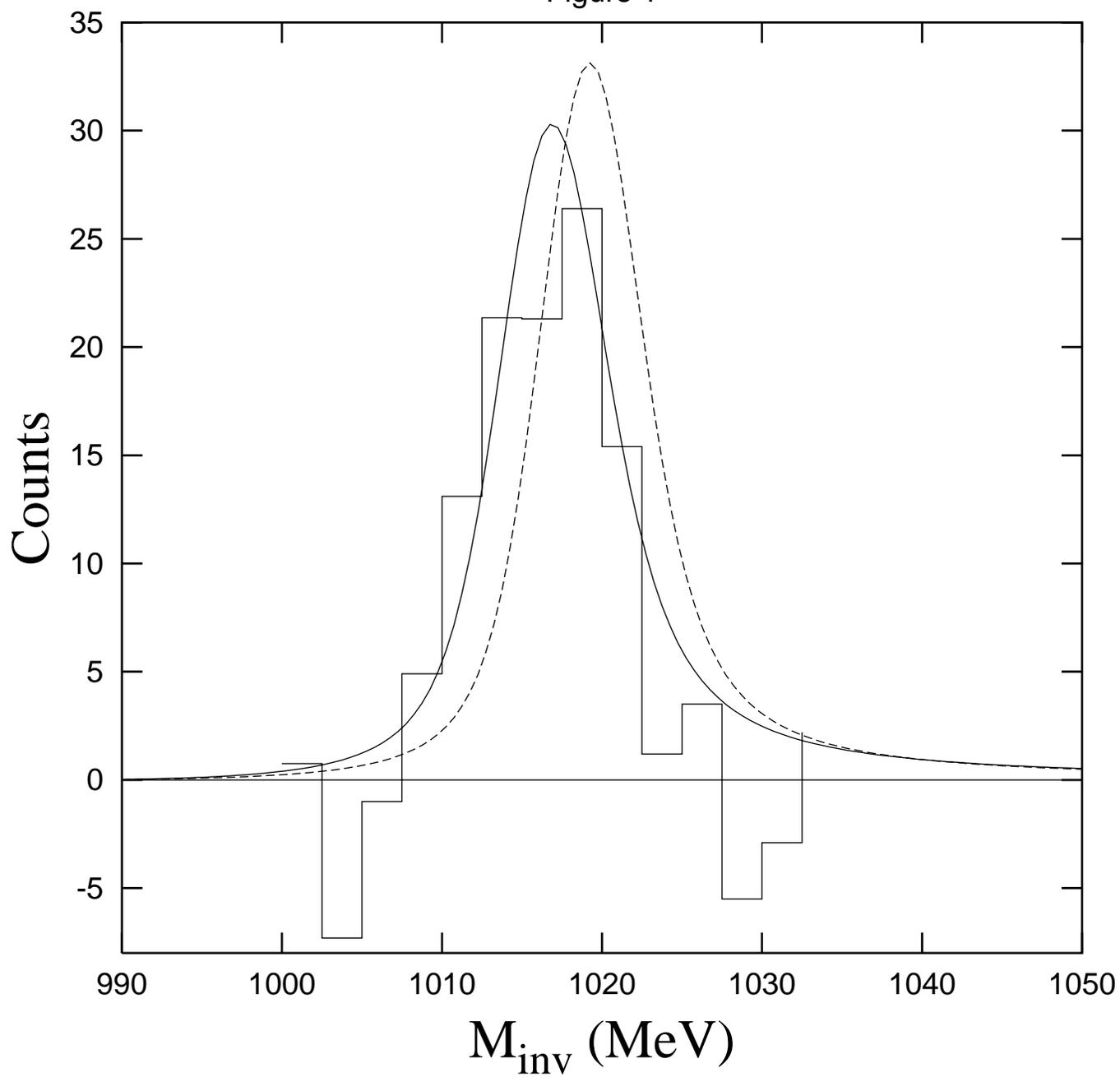
\begin{figure} 
\caption
{The background-subtracted invariant-mass spectrum for the $K^+K^-$
pairs. The histogram represents preliminary experimental data from
[5].
The dashed and solid curves represent single BW resonance terms
convoluted with a Gaussian as explained in the text. The dashed curve
corresponds to the situation where there is no shift in the $\phi$-meson
mass and width. The solid curve corresponds to $\Delta M = 2.3$ MeV
and $\Delta \Gamma = 0.78$ MeV.}
\label{1}
\end{figure}

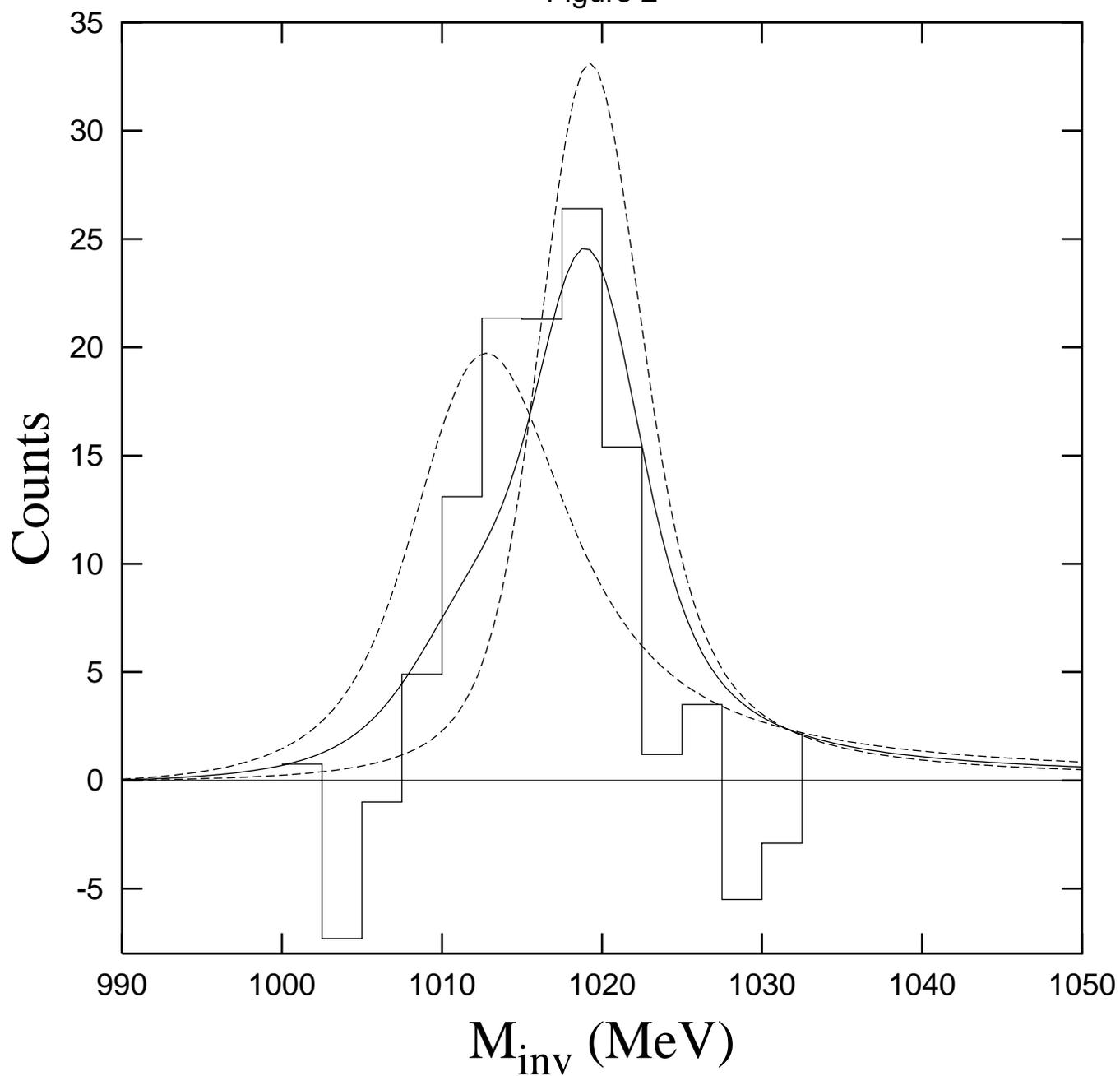
\begin{figure} 
\caption
{The background-subtracted invariant-mass spectrum for the $K^+K^-$
pairs. Experimental data as in Fig. 1. The two dashed curves
correspond to the two BW terms, one unshifted and the other
shifted. The solid curve corresponds to their sum weighted by the
factors $(1 - f)$ and $f$, respectively; see Eqs. (2-6).}
\label{2}
\end{figure}

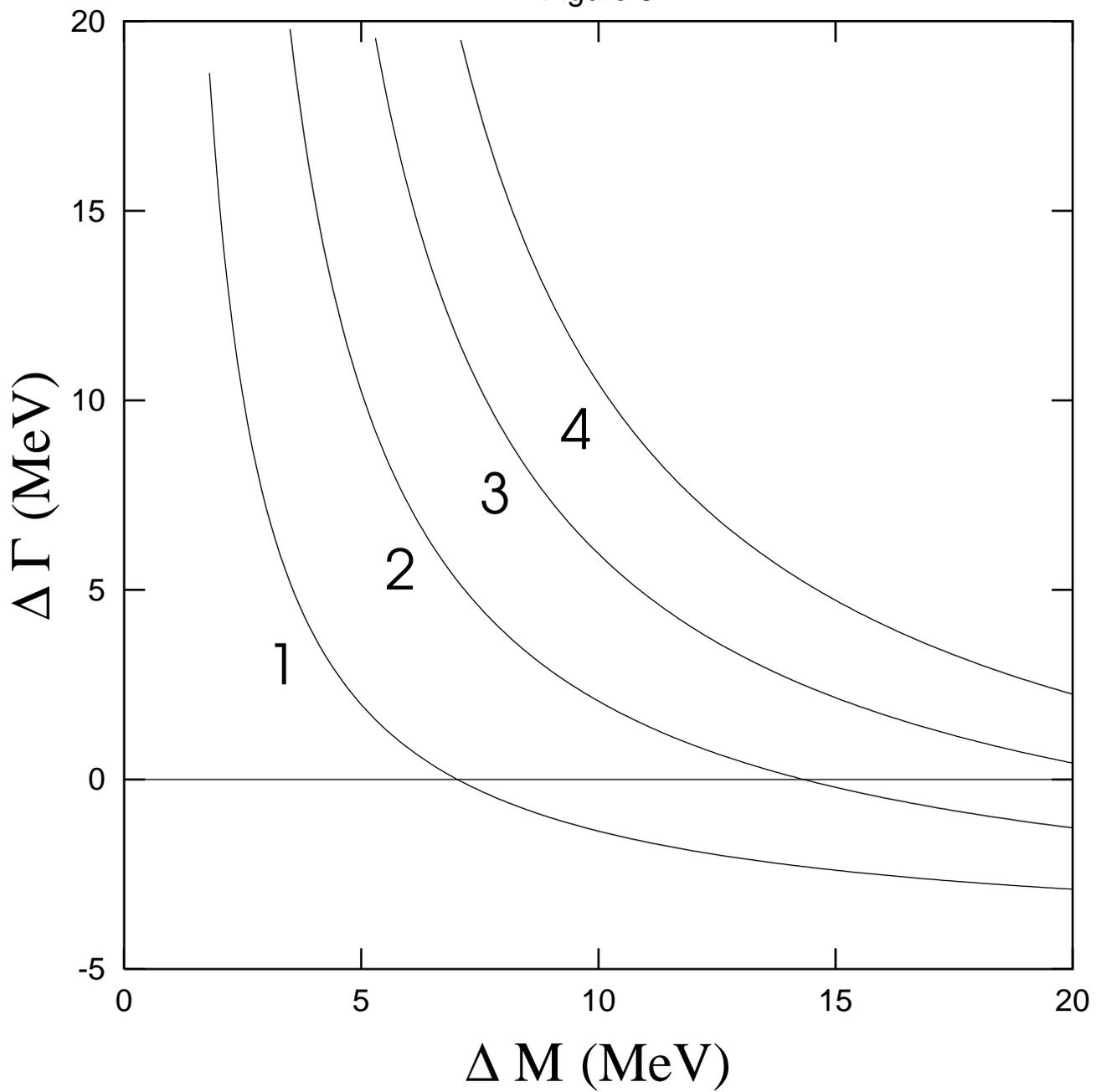
\begin{figure} 
\caption
{For a fixed $\Delta \overline M$, Eq. (10) provides a relation between
$\Delta \Gamma$ and $\Delta M$. This relation is
represented by the solid lines labeled by the value of $\Delta
\overline M$ in MeV.}
\label{3}
\end{figure}

\end{document}